\documentclass[conference]{IEEEtran}
\usepackage{csquotes}
\usepackage{cite}

\usepackage{graphicx}
\graphicspath{ {media/} }
\usepackage{float}
\usepackage{subcaption}
\usepackage{graphicx,bm}

\usepackage{amsmath}
\usepackage{bm}
\usepackage{relsize}

\usepackage{amssymb}
\usepackage{array}
\newcommand{\comm}[1]{}

\author{\IEEEauthorblockN{Vikram~Kumar\textsuperscript{*}\textsuperscript{\textdagger}, Reza~Arablouei\textsuperscript{\textdagger}, Raja~Jurdak\textsuperscript{\textdagger}\textsuperscript{*}, Branislav~Kusy\textsuperscript{\textdagger}\textsuperscript{*}, and Neil~W.~Bergmann\textsuperscript{*}}
\IEEEauthorblockA{\textsuperscript{*}School of ITEE, University of Queensland, St Lucia QLD 4072, Australia\\
\textsuperscript{\textdagger}CSIRO's Data61, Pullenvale QLD 4069, Australia}}

\begin{document}
\title{Multi-mode Tracking of a Group of Mobile Agents}
%\title{Multi-mode Tracking of Mobile Agents}

\maketitle 
\begin{abstract}
We consider the problem of tracking a group of mobile nodes, which have limited computational and energy resources, using noisy RSSI measurements and position estimates available within the group.
% The multilateration solutions are known for energy efficiency. However, these solutions are not directly applicable to dynamic grouping scenarios.
Existing solutions such as cluster-based GPS duty-cycling, individual tracking, and multilateration-based localization and tracking can only partially deal with the challenges of dynamic grouping scenarios where neighbourhoods and resource availability may frequently change. To efficiently cope with these challenges, we propose a new group-based multi-mode tracking algorithm. The proposed algorithm takes the group size and resource availability into consideration and determines the best solution at any particular time instance. We consider a clustering approach where a cluster head assigns the task of GPS activation and coordinates the usage of resources among the cluster members. We evaluate the energy-accuracy trade-off of the proposed algorithm for various fixed sampling intervals. The evaluation is based on the 2D position tracks of 40 nodes simulated using Reynolds' flocking model. For a given energy budget, the proposed algorithm reduces the mean tracking error by up to $20\%$ in comparison with the existing energy-efficient cooperative algorithms. Moreover, the proposed algorithm is as accurate as the individual-based tracking while using around 50\% less energy.   

\end{abstract}


\IEEEpeerreviewmaketitle
\section{Introduction}
Tracking has a significant role in understanding human mobility patterns, wild-life monitoring, and mobile asset tracking \cite{Allen2015, jurdak2013camazotz}. These applications involve long-term tracking of mobile agents where there are restrictions on weight and size of the tracking devices in order to minimize any disruption to the natural movement of the tracked device/individual. These restrictions limit the computational power and energy resources available to the tracking devices. The conflicting nature of long-term tracking and resource limitations highlights the need for energy-efficient tracking algorithms.

The global positioning system (GPS) has revolutionized the outdoor tracking. However, it suffers from high energy consumption and poor performance in urban areas and dense forests \cite{Vallina-Rodriguez2013}. There are significant efforts by the research community to alleviate the dominance of energy consumption by the GPS in tracking. Some propose to use inertial sensors to augment the GPS in between GPS sampling intervals. These appraoches are generally based on techniques such as extended Kalman filter, particle filter, and look-ahead filters \cite{parker2007cooperative,mohammadabadi2014cooperative}. In tracking using inertial-sensors, the accuracy is a function of the inertial sensor's sampling frequency \cite{jurdak2012group}. Tracking based on inertial sensors generally requires an occasional reliable position estimate from an independent positioning system to curb the accumulation of errors over time. 

Another approach is to use cooperation among co-located devices to share the tracking load \cite{vukadinovic2012performance,lee2012comon,jurdak2012group}. The core idea of this approach is to reduce the energy usage of a group of nodes by limiting the use of each node's GPS. The mobile nodes exploit an opportunistic grouping behavior and request for position estimates from their neighbours. A node activates its own GPS only in case position updates from its neighbouring nodes are unavailable.
%These approaches offer significant energy saving but their performance suffers from possible perturbations in the neighbouring nodes' position information as well as the radio signal due to the propagation environment. Particularly, their tracking error can be large with high levels of perturbation in the received signal strength indicator (RSSI) measurements or neighbouring nodes' position information.

Multilateration-based localization techniques are popular for their energy efficiency and simplicity. However, they may suffer from inaccuracy when there are perturbations in the position information of the anchor nodes (the nodes that have estimates of their current locations, e.g., using GPS) or in the distance estimates, e.g., using RSSI measurements between the anchor nodes and the blind node (the node interested in estimating its own position). In \cite{Tarrio2011}, the authors propose a localization algorithm based on weighted least-squares (WLS), referred to as \enquote{WLSR} hereafter, that accounts for perturbations in the RSSI measurements only. Our algorithm proposed in \cite{bias2017}, referred to as \enquote{WLSRP} hereafter, improves the WLSR algorithm by accounting for perturbation in the anchor position information as well. In contrast to the iterative and computationally complex solutions such as those based on second-order cone programming (SOCP) or semi-definite programming (SDP) \cite{angjelichinoski2014spear,Naddafzadeh-Shirazi2014,GDIEEE}, WLSR and WLSRP are closed-form and easy to implement making them suitable for localization in resource-constrained applications.

In static or slow-varying sensor networks, tracking based on multilateration is relatively straightforward. However, it is not directly applicable to long-term tracking of dynamic groups of resource-constrained mobile nodes such as humans, animals, and vehicles. The dynamic group structure brings about the challenges of frequently changing neighbouring nodes and available resources for localization. In some instances, the number of neighbouring nodes may not be sufficient to run multilateration. To address these challenges, in this paper, we propose a multi-mode group-based tracking algorithm. The proposed algorithm has three modes, namely, multilateration, cluster-based, and standalone. At any particular time instance, the proposed algorithm selects the best mode given the group structure and available resources. We use a clustering approach where a cluster head (CH) is responsible for the position update process and other activities related to cluster management.
%In some scenarios, the statistical properties of the anchor node position perturbation may not be known. Therefore, we evaluate two variants of the proposed group-based multi-mode algorithm namely \enquote{Multi-mode WLSR} and \enquote{Multi-mode WLSRP} based on the WLSR and WLSRP solutions. 

We examine the energy-accuracy trade-off offered by the proposed algorithm on the 2D position tracks of 40 nodes simulated using Reynolds' flocking model \cite{reynolds1987flocks}. The proposed algorithm provides significant improvements over the cluster-based tracking algorithm proposed in \cite{cluster2016} as well as a cooperative localization algorithm that assumes the blind node position to be the same as its nearest neighbouring node \cite{vukadinovic2012performance}. In addition, the proposed algorithm is as accurate as an individual-based tracking, where every node frequently uses its own GPS without any cooperation with other nodes, while using almost half the energy. %We also compare the results with a tracking algorithm when a  for a position update and does not involve in any cooperation with other nodes.    

\section{Related Work}

In \cite{vukadinovic2012performance} a cooperative GPS duty-cycling framework is proposed that uses a Wi-Fi ad-hoc network while ensuring application-specific error bounds. In this framework, each node tracks its own position uncertainty and whenever a node approaches an uncertainty limit, it requests a position update from other co-located nodes. In case of a fruitful reply, the node updates its position with the received information, otherwise activates its own GPS for a position update. The authors evaluate the framework on the real data traces of visitors in Epcot theme park in Florida, USA, and report some energy efficiency improvement. In \cite{jurdak2010adaptive,jurdak2012group}, the authors propose another GPS duty-cycling framework based on radio contact logging and inertial sensors. They use short-range radio communication as a measure of reducing position uncertainty of a node. If the sum of the neighbouring node position uncertainty and distance to that node is less than the node's own position uncertainty, the node updates its position estimate by neighbour's position, otherwise activates its own GPS. The algorithm also uses accelerometers of the nodes to detect their motion versus non-motion states. This avoids activation of GPS in the non-motion state resulting in better energy efficiency. The authors evaluate the performance of their algorithm on the empirical data traces of cattle. In our previous work \cite{cluster2016}, we exploit the existence of groups and propose a cluster-based tracking (CBT) of co-located mobile sensors. In this algorithm, the CH is responsible for cluster maintenance and scheduling the GPS activation at various cluster members. The CBT leads to a significant energy saving but limits the level of minimum achievable error for a given sampling interval. In this paper, we build on the concept of CBT and use an RSSI-based multilateration method as the underlying localization technique to maximize the benefit of neighborhood-based resources. 

In summary, most of the existing tracking algorithms either focus on improving tracking accuracy at the cost of high resource requirements or take energy saving as the priority and compromise accuracy. Our focus is to maximize the energy-accuracy benefit. The proposed group-based multi-mode tracking algorithm considers the available resources in a group at a particular instance of localization and then decides the best approach for localization at that instance.  
\section{Background Information}

\subsection{Problem and Assumptions}
We consider a group-based tracking problem in a 2D plane given perturbed anchor node positions and RSSI measurements. We assume additive independent zero-mean Gaussian noise ${n}_{x_i}$ and ${n}_{y_i}$ in anchor node position coordinates $\tilde{x}_i$ and $\tilde{y}_i$, $i=1,...,N$ \cite{hemmes2009cooperative}. The positive integer $N$ is the number of anchor nodes arbitrarily distributed within the communication range of the blind node. The standard deviation of noise in the anchor node position, denoted by $\sigma_{a_{i}}$, may differ among anchor nodes but is considered the same for both $x$ and $y$ axes of any particular node, i.e., 
\begin{equation}\label{pos1}
\tilde{x}_i = x_{i} + {n}_{x_i}
\end{equation} 
\begin{equation}\label{pos2}
\tilde{y}_i = y_{i} + {n}_{y_i}
\end{equation} 
\begin{equation}\label{pos3}
~{n}_{x_i}, {n}_{y_i} \sim \mathcal{N}(0,\sigma_{a_{i}}).
\end{equation} 
Here $x_{i}$ and $y_{i}$ are the original (unperturbed) position coordinates of the $i$th node. 

We consider the log-normal shadowing model for the radio signal path loss \cite{garg2001wireless}. The unperturbed RSSI measurement at the blind node for the signal transmitted from the $i$th anchor node is denoted by $p_{i}$, in the logarithmic (dBm) domain. The symbol $\tilde{p}_{i}$ denotes the perturbed version of $p_{i}$. The perturbation $n_{p_{i}}$ in $\tilde{p}_{i}$ is additive Gaussian with mean zero and standard deviation $\sigma_{p_{i}}$ (dB), i.e.,
\begin{equation}\label{pidbm1}
\tilde{p}_{i} =  p_{i} + n_{p_{i}}
\end{equation}

\begin{equation}\label{pidbmn1}
n_{p_{i}}  \sim \mathcal{N}(0,\sigma_{p_{i}}).
\end{equation}

The shadowing path-loss model describes the relationship between the RSSI measurement $p_i$ and the distance between the blind node (with coordinates $x_b$ and $y_b$) and the $i$th anchor node (with coordinates $x_i$ and $y_i$), i.e.,
\begin{equation*}
d_i=\sqrt{\left(x_i-x_b\right)^2+\left(y_i-y_b\right)^2},
\end{equation*}
as
\begin{equation}\label{pidbm2}
p_{i}=p_{0}-10\eta\log_{10}{\frac{d_i}{d_0}}
\end{equation}
where $d_0$, $p_{0}$ and $\eta$ are the reference distance, the received power at the reference distance, and the path loss exponent. Therefore, given the perturbed RSSI measurement $\tilde{p}_{i}$, the RSSI-based estimate for the distance between the blind node and the $i$th anchor node, denoted by $\tilde{d}_{i}$, is calculated as 
\begin{equation}\label{d_i_p_i_relation}
\tilde{d}_i = d_0 10^{\cfrac{\tilde{p}_{i}-p_{0}}{10\eta}}.
\end{equation}
The values of the path-loss model parameters used in our experiments are $d_0=1$ m, $p_{0}=-33.44$ dBm, and $\eta =3.567$. These values are based on the results reported in \cite{ahmad2015experiments}.% In practice,  path-loss exponent can be estimated dynamically as well \cite{mao2007path}.

\subsection{The WLSR solution}

In \cite{Tarrio2011}, the authors consider having perturbations only in the RSSI measurements and assume perfect knowledge of the anchor positions. The multilateration solution proposed in \cite{Tarrio2011} is expressed as
\begin{equation*}\label{WLSR}
\hat{\mathbf{w}}=\frac{1}{2}\left(\tilde{\mathbf{A}}^T\mathbf{S}^{-1}\tilde{\mathbf{A}}\right)^{-1}\tilde{\mathbf{A}}^T\mathbf{S}^{-1}\tilde{\mathbf{b}}
\end{equation*}
where
\begin{equation*}
\hat{\mathbf{w}} = \begin{bmatrix}\hat{x}\\\hat{y}\end{bmatrix}
\end{equation*}

\begin{equation*}
\tilde{\mathbf{A}} = \begin{bmatrix}
\tilde{x}_2 - \tilde{x}_1 & \tilde{y}_2 - \tilde{y}_1 \\
\tilde{x}_3 - \tilde{x}_1 & \tilde{y}_3 - \tilde{y}_1 \\
       ... & ... \\
\tilde{x}_N - \tilde{x}_1 &  \tilde{y}_N - \tilde{y}_1  
\end{bmatrix}
\end{equation*}

\begin{equation*}
\tilde{\mathbf{b}} = \begin{bmatrix}
\tilde{d}_1^2  - \tilde{d}_2^2 + \tilde{k}_2 - \tilde{k}_1 \\
\tilde{d}_1^2  - \tilde{d}_3^2 + \tilde{k}_3 - \tilde{k}_1\\
...\\
\tilde{d}_1^2  - \tilde{d}_N^2 + \tilde{k}_N - \tilde{k}_1
\end{bmatrix}
\end{equation*}

\begin{equation*}
\tilde{k}_i = \tilde{x}_i^2 + \tilde{y}_i^2
\end{equation*}
and the $(i,j)$th entry of the $(N-1)\times(N-1)$ matrix $\mathbf{S}$ (the covariance matrix of $\tilde{\mathbf{b}}$) is given by
\begin{equation*}\label{WLSRS}
s_{ij} = \begin{cases}
\mathrm{Var}(\tilde{d}_1^2  - \tilde{d}_{i+1}^2)  & \text{if} \ i=j\\
\mathrm{Var}(\tilde{d}_1^2) & \text{if} \ i\neq j.
\end{cases}
\end{equation*}

\subsection{The WLSRP solution}

In \cite{bias2017}, we took into account the perturbations in the anchor node position information as well and estimated and compensated for the bias induced by the non-additive nature of perturbation in $\tilde{d}_i$ owing to the adoption of the shadowing path loss model. Our proposed WLS-based bias-compensated solution (WLSRP) is given by
\begin{equation*}\label{proposed}
\hat{\mathbf{w}} = \frac{1}{2}\left(\tilde{\mathbf{A}}^T \mathbf{S}^{-1} \tilde{\mathbf{A}}\right)^{-1}\tilde{\mathbf{A}}^T\mathbf{S}^{-1}\left(\tilde{\mathbf{b}} - \textbf{c}\right)
\end{equation*}
where the $i$th entry of the vector $\mathbf{c}$ is calculated as
\begin{equation*}
c_i = \left(u^2\sigma_{p_i}^2 + \frac{u^4}{2}\sigma_{p_i}^4 \right)\left(d_1^2 - d_i^2\right) + 2\left(\sigma_{a_i}^2 - \sigma_{a_1}^2\right)
\end{equation*}
with
\begin{equation*}
u= \frac{\ln10}{5\sqrt{2}\eta}
\end{equation*}  
and the $(i,j)$th entry of $\mathbf{S}$ is computed as
\begin{equation*}\label{S_ij1}
s_{ij} = \begin{cases}
\mathrm{Var}(\tilde{d}_1^2  - \tilde{d}_{i+1}^2 + \tilde{k}_{i+1} - \tilde{k}_1) & \text{if} \ i=j\\
\mathrm{Var}(\tilde{d}_1^2 - \tilde{k}_1) & \text{if} \ i\neq j.
\end{cases}
\end{equation*}

\section{Proposed Algorithm}

The description of the proposed multi-mode group-based tracking algorithm is as follows:

\subsection{Cluster formation and CH selection}

The first task is to form  clusters among a group of nodes based on a predefined communication range. For this purpose, all the nodes initially acquire their position information from an independent positioning source, e.g., GPS. Then, the position information is shared within the communication range. To decide the CH, each node runs a random timer and the node with the least value of the timer is selected as the CH \cite{xu2002heterogeneous}. The ties in the timer value are resolved on the basis of lowest node identifier value. A cluster member may have connectivity to only a few of the cluster members but all members have assured connectivity to the CH. Each node can belong to one cluster only and selection among multiple clusters is made on the basis of the distance to the CH estimated using RSSI measurements.

\subsection{Cluster merging and node splitting}

There can be multiple clusters with varying numbers of nodes in each cluster at any given instance of time. Both the number of clusters and their members change over time. For the situation where two or more CHs come within the communication range of each other, a cluster merging is performed. The cluster merging process follows the rule of \emph{the bigger cluster absorbs the smaller cluster}.

A  node leaves its current cluster and seeks to join a new one if it does not receive three scheduled updates from its current CH. In case there is no nearby CH, the node starts a new cluster formation process.

\subsection{Position update process}

The CH is responsible for all the cluster management activities. It is also responsible for providing support in position update activity to the cluster members. Whenever a position update is required, CH checks the number of members in its cluster and their energy levels to choose one of the following modes accounting for the temporary grouping behavior of the nodes.

\subsubsection{Multilateration mode}

The algorithm enters this mode if the cluster size $C_s$ is more than a threshold $C_t$. In this mode, CH randomly selects $A$ number of nodes as the anchor nodes. The values of $C_t$ and $A$ are decided based on the cost-benefit analysis for the nodes. A combination of a big value of $C_t$ and a low value of $A$ means higher energy efficiency but lower positioning accuracy. The anchor nodes update their positions through GPS and share it among the cluster members. Then, all the cluster members use an RSSI-based multilateration technique (WLSR or WLSRP) to estimate their current positions. Here, we assume that changes in the cluster geometry are negligible while taking the GPS and the subsequent RSSI  measurements. A cluster member discards a position estimate if the distance between any anchor node and the estimated position is greater than double the communication range. The node then assumes its position to be the same as that of the nearest anchor node selected to have the smallest RSSI-based distance. This helps the nodes to filter out the inaccurate position estimates.% given by the RSSI-based multilateration technique.

\subsubsection{Cluster-based mode}

The algorithm enters this mode, if $1<C_s\leq C_t$. In this mode, CH determines the nodes that have the minimum energy required for GPS sampling and randomly chooses one of them to be the anchor node. Every cluster member assumes their position to be the same as that of this anchor node. In the absence of new GPS position updates either from the GPS or from any neighbor, the node uses its last position as the current position estimate. 

\subsubsection{Standalone mode}

The algorithm enters this mode if $C_t=1$ or if a particular node does not receive three consecutive updates while being part of a group in the cluster-based or multilateration modes. In this mode, the node activates its own GPS for the current position update and later on searches for an existing cluster. In case of no nearby existing cluster, the node starts the process of forming a new cluster. 
\begin{figure}
  \begin{subfigure}[b]{0.248\textwidth}
    \includegraphics[width=\textwidth]{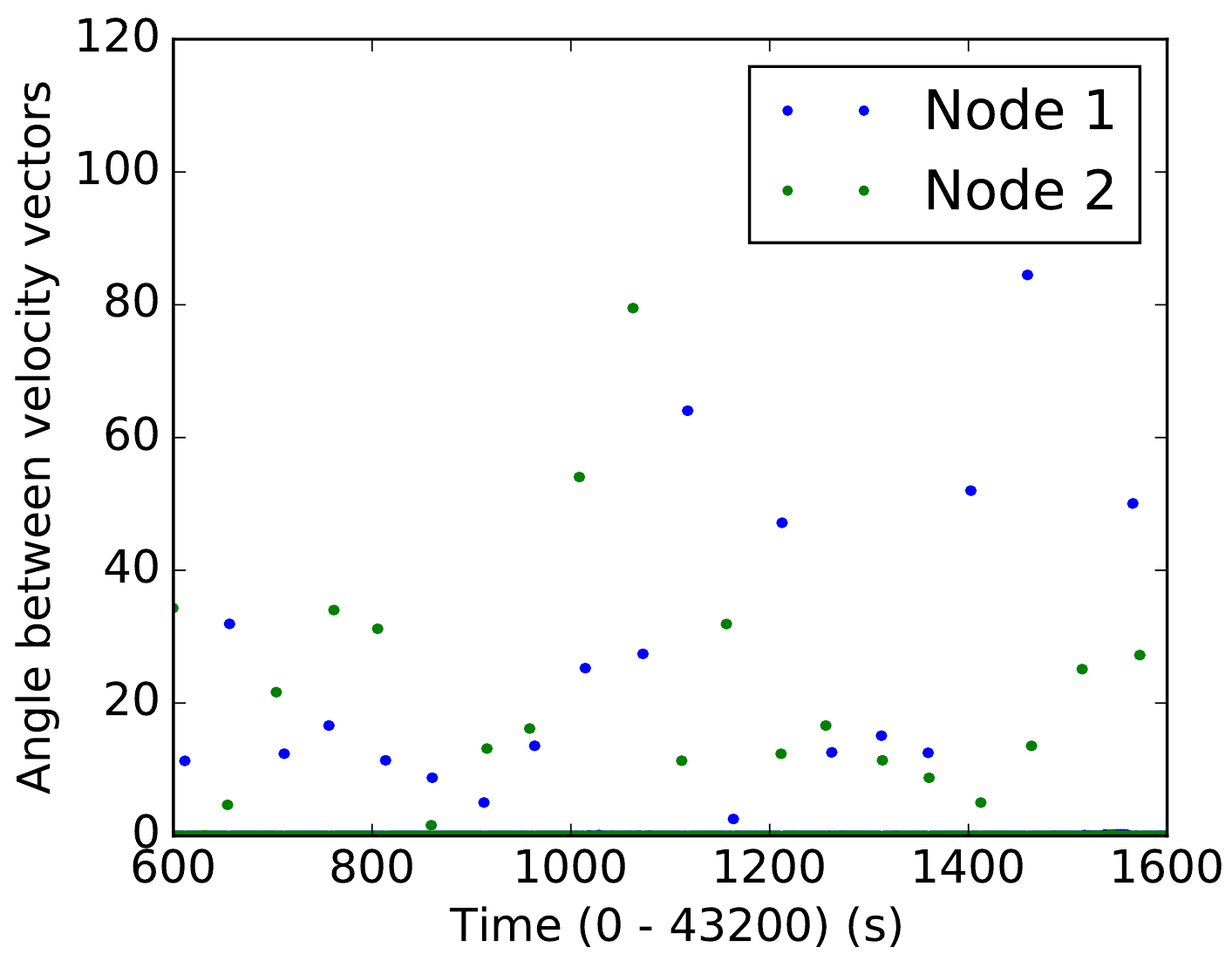}
    \caption{}
    \label{fig:b}
  \end{subfigure}
  \begin{subfigure}[b]{0.23\textwidth}
    \includegraphics[width=\textwidth]{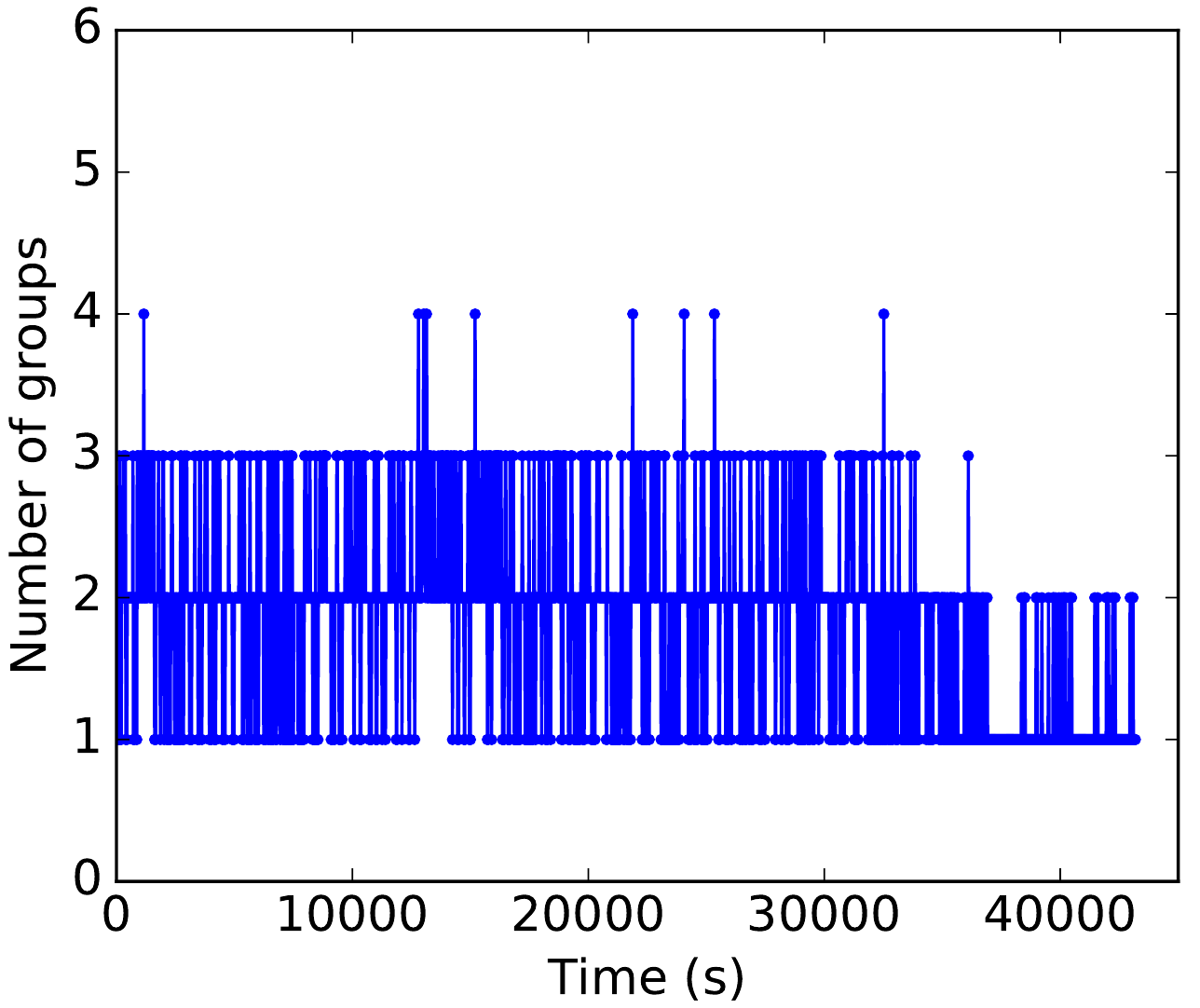}
    \caption{}
    \label{fig:d}
  \end{subfigure}
  \caption{Movement data insights. (a) A time-slice view of the angle between two consecutive velocity vectors of a node. Variation in the angle implies that the track followed by the node is not a straight line. (b) Numbers of the clusters formed for a particular position sampling interval over the entire simulation period. The variation shows the dynamic clustering nature of the simulated data.}
  \label{fig:D1}
\hspace*{\fill}
\end{figure}

\section{Experimental Setup}

\subsection{Movement data}

We design a python-based simulator to generate group movements of $40$ mobile nodes, e.g., bats, in an area of $50$km$\times50$km. The scenario consists of two living areas and one foraging area. All nodes start their journey from the living areas towards the foraging area at a maximum speed of $6$m/s while aiming to maintain an inter-node distance of approximately $20$m. We generate a total of $43200$ positions per node at the time resolution of one second. We use Reynolds' flocking model to generate the positions during the journey \cite{reynolds1987flocks}. To generate movements in the foraging area, we use a random walk movement model. During the journey, there are formations of different clusters among the nodes. Both cluster number and their member counts vary in time. Some more information about the data generated through the considered simulation framework is given in Fig. \ref{fig:D1}.

\subsection{Energy model}

The assumed energy model is based on a multi-mode mobile sensing device known as Camazotz \cite{jurdak2013camazotz}. The device consists of a CC430 system on chip with GPS, inertial, temperature, acoustic, and air-pressure sensors. It has two solar panels and a $300$mAh Li-Ion battery operating at $3.7$V ($3996$J). The total weight of the device is under $30$g making it suitable for tracking wildlife and mobile industrial assets.

We identify the major energy consuming activities and calculate their energy usage based on the information given in the data sheets and reported in \cite{jurdak2013camazotz}. To simplify the energy model, we assume that the GPS only operates in hot start and the energy consumption is the same for data transmission and reception activities. We consider the total energy that nodes consume at standby mode as well as for cluster management as miscellaneous energy and focus our analysis on the activities that dominate the energy usage, i.e., sensor sampling and radio communication. We calculate the total energy consumed by the GPS sampling activity, denoted by $E_g$, and the total energy required for a data transmission or reception activity, denoted by $E_r$, via
\begin{equation}\label{Eg}
    E_g = T_g\left( P_g + P_m\right)
\end{equation}
\begin{equation}\label{Er}
    E_r = T_t\left( P_r + P_m\right)
\end{equation}
where the values of the parameters are given in Table \ref{tab:ED}.

\begin{table}[t]
    \centering
    \caption{Details of energy model}
    \label{tab:ED}
    \begin{tabular}{cc}%{| m{6cm}| m{1cm} |} 
        \hline
        \textbf{Activity / Component}   & \textbf{Value}  \\
        \hline
         Total simulation period $(T)$  & $43200$s  \\ 
        \hline
         GPS power consumption $(P_g)$  & $74$mW  \\ 
        \hline
       GPS activity time hot-start mode   $(T_g)$ & $5$s \\ 
        \hline
         MCU power consumption $(P_m)$ & $13.2$mW \\ 
        \hline
          Radio power consumption $(P_r)$ & $13.2$mW \\ 
        \hline
         Packet size $(S)$ & $80$bits \\ 
        \hline
          Channel bit rate  $(C)$ & $256$Kbps \\ 
        \hline
          Packet transmission/receiving time $(S/C)$ $(T_t)$ & $0.31$ms \\ 
        \hline
            Standby power  $(P_s)$ & $1.2$$\mu W$ \\ 
        \hline 
          Miscellaneous energy consumption $(E_l)$ & $ 54$ J \\ 
        \hline
    \end{tabular}
\end{table}

%\begin{table}[t]
%    \centering
%    \caption{Activity-wise energy consumption}
%    \label{tab:AD}
%    \begin{tabular}{| m{4cm}| m{2cm} |} 
%        \hline
%        \textbf{Activity}   & \textbf{Energy/activity}  \\
%        \hline
%         GPS  $(E_g)$  & $0.436$J  \\ 
%        \hline
%         Radio communication $(E_r)$  & $34.78$ $\mu J$  \\ 
%        \hline
%    \end{tabular}
%\end{table}
\begin{figure*}
    \centering   
    \includegraphics[width=\textwidth]{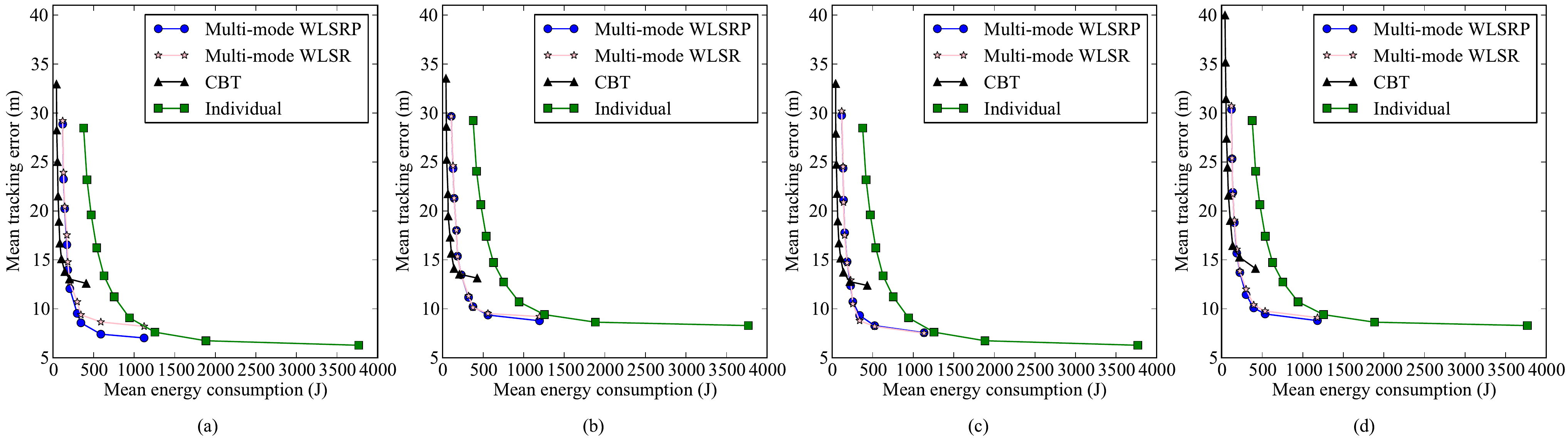}  
    \caption{Performance of the simulated algorithms in different perturbation scenarios. The top-left and bottom-right points of the curves correspond to sampling intervals of $50$s and $5$s, respectively, and the points in between to the intermediate sampling intervals of $5$s apart. (a) $\sigma_{p_i} =1$dB and $\sigma_{a_i} = 1,5$m; (b) $\sigma_{p_i} =1$dB and $\sigma_{a_i} = 5,10$m; (c) $\sigma_{p_i} =3$dB and $\sigma_{a_i} = 1,5$m; (d) $\sigma_{p_i} =3\rm{dB}$ and $\sigma_{a_i} = 5,10$m.}
    \label{fig:results1}
\end{figure*}

\subsection{Experiment details}

We simulate four scenarios differentiated based on the amount of perturbation in the GPS-based positions and RSSI measurements. In all four scenarios, $50\%$ of the nodes have a high GPS performance (small values of $\sigma_{a_i}$) and $50\%$ have a low GPS performance (large values of $\sigma_{a_i}$). Therefore, at any particular movement, a cluster can have a mixture of devices with low and high GPS performance. We set the value of $\sigma_{a_i}$ in the range of $1$m to $10$m. The clustering performance highly depends on the accuracy of the underlying ranging technique. Hence, we evaluate the performance of the proposed algorithm for values of $\sigma_{p_i} = 1$dB and  $\sigma_{p_i} = 3$dB. We use a clustering threshold of $C_t = 10$ and set the number of anchor nodes for all instances of localization to $N=6$. This value is based on the results of our preliminary experiments to find the number of anchor nodes required to achieve a reasonable mean localization error. The value of $C_t$ is based on a trade-off between $N$ and the average cluster size in the simulated data.

\section{Simulation Results}

We consider the mean tracking error and mean energy consumption of all nodes for any given sampling interval as the performance evaluation criteria. To calculate the mean tracking error, first, we perform a linear interpolation between any two consecutive points of the sampled trajectory of a node to match the maximum available resolution of $1$s in the original trajectories. Then, we calculate each node's individual tracking error as the mean of the Euclidean distances between the points of the trajectory obtained after linear interpolation and their corresponding points of the original trajectory. Finally, the mean tracking error is calculated as the mean of all node's individual tracking errors. Similarly, the energy consumption of each node is calculated based on its individual activities during tracking through the energy model given by \eqref{Eg} and \eqref{Er}. We calculate the mean energy consumption by taking the average of energy consumed by all nodes.

The results are presented in Fig. \ref{fig:results1} for two variants of the proposed algorithm called multi-mode WLSR and multi-mode WLSRP, which, respectively, use WLSR and WLSRP solutions for multilateration. The tracking algorithms based on solely multilateration (WLSR or WLSRP) had large errors due to lack of sufficient number of anchor nodes at some instances of localization; hence, we excluded them. \enquote{CBT} is the cluster-based tracking used in \cite{cluster2016,vukadinovic2012performance,jurdak2012group}. In CBT, the blind node assumes its position to be the same as the position estimate of its neighboring node with the latest position information. \enquote{Individual} refers to a tracking algorithm based on standalone GPS sampling. The group tracking performance is evaluated over sampling intervals ranging from $5$s to $50$s. The top-left and bottom-right points of each curve in Fig. \ref{fig:results1} show the results for a sampling interval of $50$s and $5$s, respectively. The points in between correspond to the intermediate intervals following the order with equal step size of $5$s. 

First, we consider the effects of various combinations of perturbation levels on the performance of evaluated group-based tracking algorithms. The individual tracking algorithm has no dependence on the RSSI measurements. Therefore, the performance of individual tracking is the same in Fig. \ref{fig:results1}a and Fig. \ref{fig:results1}c and similarly in Fig. \ref{fig:results1}b and  Fig.\ref{fig:results1}d. The effect of the increase in the standard deviation of noise in anchor position $\sigma_{a_i}$ on the performance of individual tracking can be seen as the increase in the error from $6$m to $9$m for low sampling intervals in Fig. \ref{fig:results1}a and Fig. \ref{fig:results1}b. It appears that linear interpolation errors dominate the localization errors in high-sampling intervals reducing the effect of an increase in the value of $\sigma_{a_i}$. All the other algorithms have dependence on the RSSI measurement errors resulting in performance change from one scenario to another. The effect on the performance of various combinations of perturbations is more noticeable at lower sampling intervals due to the dominance of the linear interpolation errors at the higher sampling intervals.

\begin{figure}
    \centering   
    \includegraphics[width=0.5\textwidth]{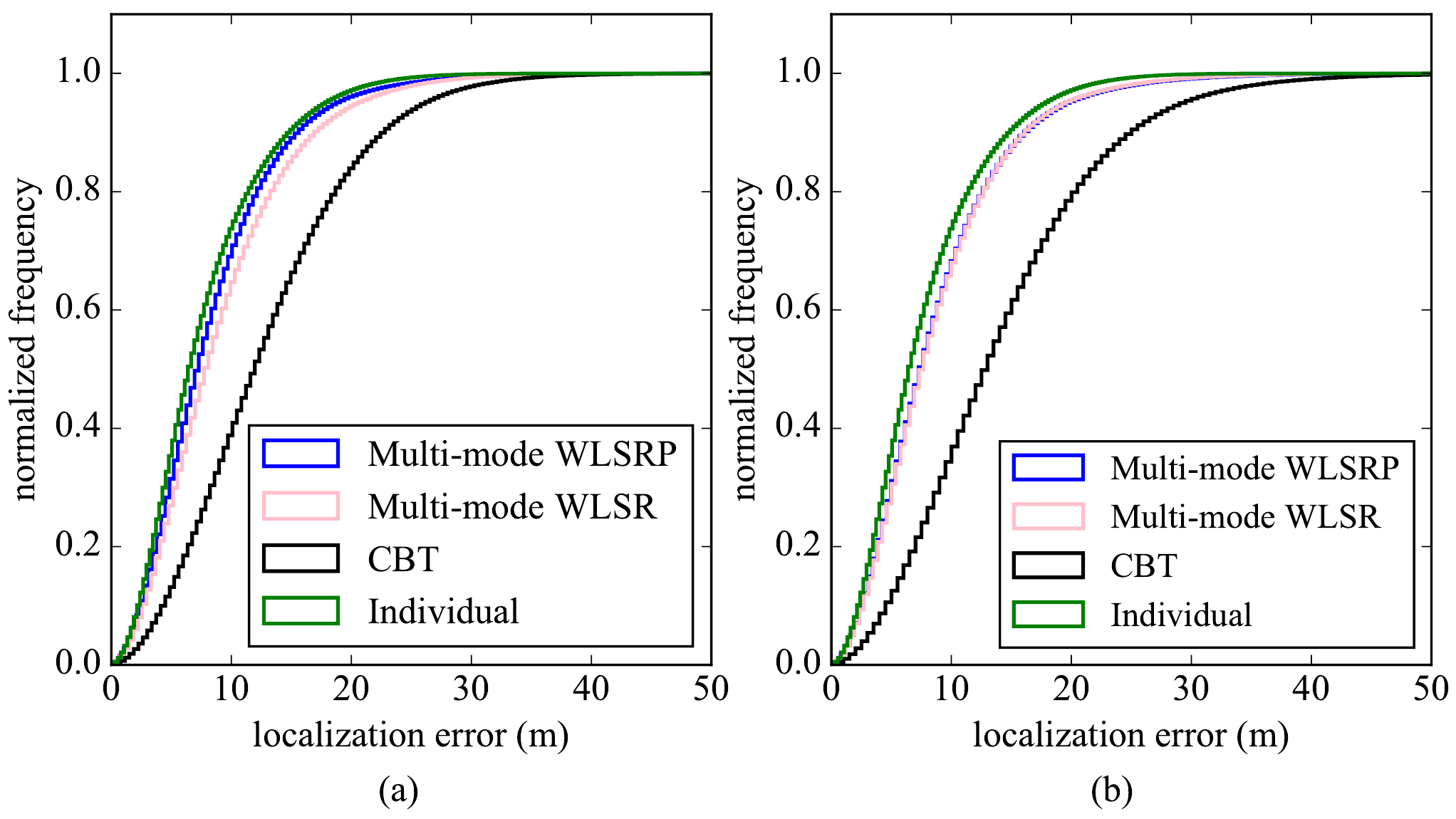}  
    \caption{Cumulative distribution of mean tracking error of the simulated algorithms with a sampling interval of $5$s, $\sigma_{a_i}=5,10$m, and (a) $\sigma_{p_i} =1$dB or (b) $\sigma_{p_i} =3$dB.}
    \label{fig:results2}
 \end{figure}

Both variants of the proposed multi-mode group-based tracking algorithm provide significant energy savings over the individual tracking while performing similarly in terms of mean tracking error. As an example, for the sampling interval of $10$s, the proposed algorithm consumes $50\%$ less energy than the individual tracking. In a point-to-point sampling-interval-wise comparison, both variants of the proposed algorithm have lower tracking error than the CBT algorithm but consume more energy with most of the sampling intervals. For low sampling intervals such as $20$s, both variants of the proposed algorithm have roughly half the tracking error while consuming similar amount of energy. 

The performance of the multi-mode WLSRP and multi-mode WLSR variants is comparable for higher sampling intervals in all the scenarios. However, multi-mode WLSRP performs better for low sampling intervals. For example, in Fig. \ref{fig:results1}a, the multi-mode WLRP has $20\%$ less tracking error as compared to the multi-mode WLSR with a sampling interval of $10$s. In general, taking into account the anchor noise perturbation (by multi-mode WLSRP) is more beneficial in low sampling intervals.

We present the cumulative distributions of mean tracking error for two arbitrary scenarios in Fig. \ref{fig:results2}. The cumulative distributions of both variants of the proposed algorithm are close to that of the individual-based tracking algorithm and attest to a significantly better performance compared with the CBT algorithm.

In summary, the proposed multi-mode group-based tracking algorithm offers substantial improvements in terms of energy efficiency and tracking accuracy over the individual as well as CBT tracking algorithms. The offered performance benefits are more prominent with high perturbation levels and low sampling intervals.

\section{Conclusion}

We proposed an energy-efficient multi-mode group-based tracking algorithm by combining efficient and accurate multilateration techniques with practical and flexible cluster processing to address the challenges of tracking a dynamic group of mobile agents. We evaluated the performance of the proposed algorithm in comparison with the existing related algorithms using a dataset generated via Reynolds' flocking algorithm and considering various amounts of perturbation in RSSI and anchor position information. The proposed algorithm was shown to preform favorably against the existing cluster-based and individual-based group tracking algorithms. The performance improvement afforded by the proposed algorithm was more pronounced when having high perturbation levels and low sampling intervals.
%Compared to multi-mode WLSR, accounting for anchor position perturbations in multi-mode WLSRP is more beneficial in low sampling intervals. The proposed algorithm delivers comparable performance to the individual while consuming half the energy. 

\section{Acknowledgement}

Vikram Kumar gratefully acknowledges the financial support received from the Government of India under the National Overseas Scholarship scheme.

\bibliographystyle{IEEEtran}
\bibliography{References} 

\end{document}